%% file: main.tex
\documentclass{article}

\usepackage{arxiv}

\usepackage[utf8]{inputenc}
\usepackage[T1]{fontenc}
\usepackage{hyperref}
\usepackage{url}
\usepackage{amsmath}
\usepackage{booktabs}
\usepackage{multirow}
\usepackage{amsfonts}
\usepackage{nicefrac}
\usepackage{microtype}
\usepackage{cleveref}
\usepackage{graphicx}
\usepackage{caption}
\usepackage{pgfplots}
\pgfplotsset{compat=1.18}
\input{figures/chart_common}
\usepackage[round]{natbib}
\usepackage{doi}

\newcommand{\benchmark}{LATTICE}

\title{\benchmark: Evaluating Decision \\ Support Utility of Crypto Agents}

\author{
	Aaron Chan$^{1}$ \hspace{0.5em} Tengfei Li$^{1\dagger}$ \hspace{0.5em} Tianyi Xiao$^{1\dagger}$ \hspace{0.5em} Angela Chen$^{1}$ \hspace{0.5em} Junyi Du$^{1}$ \hspace{1em} Xiang Ren$^{1,2}$ \vspace{1em} \\
	$^{1}$Sahara AI \hspace{1em} $^{2}$University of Southern California \vspace{0.5em} \\
	{\small $^{\dagger}$Equal contribution.} \vspace{0.5em} \\
	\texttt{aaronc@saharalabs.ai}
}

\renewcommand{\headeright}{Technical Report}
\renewcommand{\undertitle}{Technical Report}
\renewcommand{\shorttitle}{\benchmark: Evaluating Decision Support Utility of Crypto Agents}

\hypersetup{
	pdftitle={\benchmark: Evaluating Decision Support Utility of Crypto Agents},
	pdfsubject={cs.AI, cs.CL},
	pdfauthor={Aaron Chan, Tengfei Li, Tianyi Xiao, Angela Chen, Junyi Du, Xiang Ren},
	pdfkeywords={LATTICE; cryptocurrency agents; decision support; benchmark; LLM evaluation; LLM-as-judge; human-in-the-loop; DeFi; conversational agents},
}

\begin{document}
\maketitle

\input{0_abstract}

\input{1_intro}

\input{2_related}

\input{3_benchmark}

\input{4_experiments}

\input{5_conclusion}

\bibliographystyle{unsrtnat}
\bibliography{references}

\end{document}

%% file: figures/chart_common.tex
\definecolor{cpElsa}{HTML}{4477AA}
\definecolor{cpGina}{HTML}{E69F00}
\definecolor{cpJune}{HTML}{332288}
\definecolor{cpMinara}{HTML}{C73E9D}
\definecolor{cpSorin}{HTML}{0F766E}
\definecolor{cpSurf}{HTML}{66CCEE}

\pgfplotsset{
	experiment bar axis/.style={
		axis lines*=left,
		axis line style={draw=black!35, line width=0.5pt},
		tick style={draw=black!35, line width=0.5pt},
		ylabel style={font=\small, yshift=2pt},
		x tick label style={font=\small, black!80},
		y tick label style={font=\small, black!55},
		ymajorgrids=true,
		major grid style={line width=0.4pt, draw=black!8},
		tick align=outside,
		scaled y ticks=false,
		clip=false,
	},
	experiment legend/.style={
		font=\small,
		legend cell align=left,
		/tikz/every even column/.append style={column sep=0.9em},
		draw=none,
		fill=none,
		inner sep=4pt,
	},
}

%% file: 0_abstract.tex
\begin{abstract}
    We introduce \benchmark, a benchmark for evaluating the \textit{decision support utility} of crypto agents in realistic user-facing scenarios.
    Prior crypto agent benchmarks mainly focus on reasoning-based or outcome-based evaluation, but do not assess agents' ability to assist user decision-making.
    \benchmark\ addresses this gap by: (1) defining six evaluation dimensions that capture key decision support properties; (2) proposing 16 task types that span the end-to-end crypto copilot workflow; and (3) using LLM judges to automatically score agent outputs based on these dimensions and tasks.
    Crucially, the dimensions and tasks are designed to be evaluable at scale using LLM judges, without relying on ground truth from expert annotators or external data sources.
    In lieu of these dependencies, \benchmark's LLM judge rubrics can be continually audited and updated given new dimensions, tasks, criteria, and human feedback, thus promoting reliable and extensible evaluation.
    While other benchmarks often compare foundation models sharing a generic agent framework, we use \benchmark\ to assess production-level agents used in actual crypto copilot products, reflecting the importance of orchestration and UI/UX design in determining agent quality.
    In this paper, we evaluate six real-world crypto copilots on 1{,}200 diverse queries and report breakdowns across dimensions, tasks, and query categories.
    Our experiments show that most of the tested copilots achieve comparable aggregate scores, but differ more significantly on dimension-level and task-level performance.
    This pattern suggests meaningful trade-offs in decision support quality: users with different priorities may be better served by different copilots than the aggregate rankings alone would indicate.
    To support reproducible research, we open-source all \benchmark\ code and data used in this paper.%
    \renewcommand{\thefootnote}{\fnsymbol{footnote}}%
    \setcounter{footnote}{0}%
    \footnote{Code and data are available at: \url{https://github.com/SaharaLabsAI/lattice-benchmark}.}%
\end{abstract}
\renewcommand{\thefootnote}{\arabic{footnote}}%
\setcounter{footnote}{0}%


%% file: 1_intro.tex
\section{Introduction}
\label{sec:intro}

\subsection{Motivation}

Crypto agents are AI assistants---usually powered by large language models (LLMs) equipped with tools and live data feeds---that help users navigate protocols, assess risk, and take action in fast-moving crypto markets.
In practice, crypto agents are commonly realized as \textit{crypto copilots}, consisting of chat-based experiences embedded in wallets, trading tools, protocol pages, etc.
These copilots operate in a \textit{human-in-the-loop} setting, where they provide information, give recommendations, and execute actions, while users are responsible for interpreting outputs and making final decisions.
In this setting, crypto agent success depends not only on reasoning capability or prediction accuracy, but also on how well agent outputs align with user intent, communicate clear explanations, and enable actionable next steps.
We collectively refer to these kinds of user-centered properties as the \textit{decision support utility} of a crypto agent, as they determine whether users can appropriately understand, assess, and act upon agent outputs.
However, there is currently no benchmark that systematically evaluates this aspect of agent quality.

Existing crypto agent benchmarks largely focus on reasoning-based or outcome-based evaluation.
\textit{Reasoning-based} benchmarks evaluate crypto agents on their ability to reason over dynamic market and on-chain data, emphasizing capabilities such as retrieval, synthesis, and tool usage~\citep{guo2025cryptobench, eswaran2026cryptoanalystbench, dai2025hallucination}.
Reasoning performance is typically assessed using intrinsic criteria such as factual correctness, logical coherence, or cross-source integration.
\textit{Outcome-based} benchmarks evaluate crypto agents on their ability to make decisions or predictions in real-world crypto market settings, emphasizing capabilities such as market timing, portfolio allocation, and strategy adaptation~\citep{li2025investorbench, qian2025agents, zeng2025futurex}.
Outcome performance is typically assessed using extrinsic criteria such as financial returns, risk-adjusted portfolio performance, or forecasting accuracy.
Both types of benchmarks generally rely on expert annotators or external data sources to construct task inputs or ground truth references.

While these approaches provide valuable insights about crypto agents' reasoning processes and realized outcomes, they do not specifically consider how crypto agents function in user-facing interactions.
In particular, existing benchmarks do not evaluate how crypto agent outputs are interpreted and contextualized within a user's decision-making process, nor how effectively they support user decision-making.
Moreover, they do not capture differences in crypto agent response structure and explanation quality, which can vary substantially even when underlying reasoning or outcomes are similar.
As a result, crypto agents that perform comparably under existing benchmarks may still differ significantly in helpfulness and trustworthiness to end users in human-in-the-loop settings.
This leaves a gap in evaluating whether crypto agents effectively support user decision-making in practice, thus motivating a complementary class of benchmarks focused on decision support utility.

\subsection{Our Approach}

We introduce \textbf{\benchmark} (\textbf{L}ayered Evalu\textbf{A}tion of Decision Suppor\textbf{T} U\textbf{T}ility \textbf{I}n \textbf{C}rypto Ag\textbf{E}nts), a benchmark for evaluating the decision support utility of crypto agents in realistic user-facing scenarios. Rather than measuring success in terms of reasoning quality or realized outcomes, \benchmark\ evaluates how effectively agent outputs support user decision-making in human-in-the-loop settings.

The \benchmark\ stack consists of three core layers: evaluation dimensions, tasks, and query categories.
\benchmark\ operationalizes decision support utility through six evaluation dimensions---Intent Fidelity, Mechanism Clarity, Uncertainty Handling, Actionability, Evidence Coverage, and Response Structure---which together capture whether a response can be understood, evaluated, and acted upon by a user.
To reflect real-world usage, \benchmark\ defines sixteen task types spanning the end-to-end crypto copilot workflow, including market research, risk analysis, execution planning, and capital allocation.
\benchmark\ also includes five query categories---Basic, Comparative, Constrained, Decision, and Ambiguous---in order to test how crypto agents respond to different core situations within a given task.

A central design goal of \benchmark\ is to enable scalable evaluation by minimizing dependencies on expert annotation or external data sources for constructing task inputs or ground-truth references.
To achieve this, \benchmark's evaluation dimensions and tasks are carefully designed such that crypto agents can be assessed by LLM judges given only the final agent response and its corresponding query.
To ground the benchmark in real-world usefulness, \benchmark's evaluation dimensions, tasks, and rubric are also developed based on substantial input from domain experts.
However, once these benchmark components are in place, \benchmark\ can be deployed and reused at scale across large sets of LLM-generated queries and agent responses.

\benchmark\ uses LLM judges to score each agent response independently using a rubric based on the six dimensions, producing structured scores and rationales.
Scores are then aggregated using query-category-dependent weights, and the evaluation protocol enforces per-agent absolute scoring rather than pairwise agent comparisons.
This rubric-driven setup enables consistent evaluation across agents while reducing sensitivity to superficial stylistic differences.
Furthermore, \benchmark's LLM judge rubric can be continually audited and updated with little additional effort, given new dimensions, tasks, criteria, or human feedback. 
This adaptability to evolving crypto agent behavior and user needs helps promote reliable and extensible evaluation.

In contrast to other benchmarks that compare foundation models sharing a generic agent framework, \benchmark\ evaluates crypto agents as they are deployed in actual crypto copilot products, capturing the combined effects of model behavior, orchestration, tool usage, and UI/UX design on decision support quality.
In this paper, we use \benchmark\ to test six well-known or emerging real-world copilots---Elsa, Gina, June, Minara, Sorin, and Surf---on 1{,}200 diverse queries, reporting performance both in aggregate and across dimensions, tasks, and query categories.

Our results show that while aggregate scores are often similar across copilots, meaningful differences emerge at the level of individual dimensions and tasks, revealing trade-offs in decision support utility that are not captured by existing benchmarks.
These differences suggest that users with different priorities may prefer different copilots, even when aggregate rankings are comparable.
In addition, we find that human pairwise preference judgments are broadly consistent with \benchmark\ rankings, indicating that the benchmark captures aspects of output quality that align with user perceptions.
To support reproducible research, we open-source all \benchmark\ code and data used in this paper.

%% file: 2_related.tex
\section{Related Work}
\label{sec:related}

\subsection{Crypto Agent Benchmarks}

A growing number of benchmarks evaluate AI agents in crypto use cases, with respect to areas like analysis, execution, adversarial robustness, and cross-market decision-making~\citep{guo2025cryptobench, eswaran2026cryptoanalystbench, dai2025hallucination, caiba2025cryptonerbenchmark, caiba2025onchainexecutionbenchmark, li2024cryptotrade, yu2024fincon, luo2025llm, li2025investorbench, qian2025agents, zeng2025futurex}.
These benchmarks largely fall into reasoning-based and outcome-based evaluation paradigms, differing primarily in the criteria used to assess agent performance.

\textbf{Reasoning-based benchmarks} evaluate agents on their ability to analyze market and protocol data, often in an analyst-style setting.
They emphasize capabilities such as retrieval, synthesis, and tool use, and are typically assessed using intrinsic criteria such as factual correctness, logical coherence, and cross-source integration~\citep{guo2025cryptobench, eswaran2026cryptoanalystbench}.
Some benchmarks probe robustness under adversarial or high-risk conditions~\citep{dai2025hallucination}, or isolate narrower operational capabilities such as entity extraction or transaction execution~\citep{caiba2025cryptonerbenchmark, caiba2025onchainexecutionbenchmark}.
System-level evaluations extend this paradigm to end-to-end workflows such as trading or research pipelines, while still primarily measuring the correctness or completeness of intermediate analyses~\citep{li2024cryptotrade, yu2024fincon, luo2025llm}.

\textbf{Outcome-based benchmarks} evaluate agents based on the realized performance of their decisions, using extrinsic criteria such as financial returns, risk-adjusted portfolio performance, or forecasting accuracy~\citep{li2025investorbench, qian2025agents, zeng2025futurex}.
These benchmarks typically embed agents in simulated or live environments involving trading, portfolio management, or cross-market decision-making, where success is defined by downstream outcomes.
Some benchmarks focus on forecasting-style tasks, evaluating predictive accuracy under time-evolving information and uncertainty~\citep{zeng2025futurex}.
Others evaluate agents in trading or market-based settings, where decisions are executed and performance is measured through realized economic outcomes such as profit or portfolio returns~\citep{li2025investorbench, qian2025agents}.

These prior crypto agent benchmarks focus on evaluating either intrinsic reasoning quality or extrinsic decision outcomes, but do not capture how agent outputs support user decision-making in interactive, human-in-the-loop settings.

\subsection{Forecasting and Prediction Market Benchmarks}

Beyond domain-specific agent benchmarks, a complementary line of work evaluates LLM agents in forecasting and prediction market settings, across domains such as economics, politics, and general financial markets~\citep{ye2024mirai, zeng2025futurex, arora2026predictionmarketbench, yu2025livetradebench, together2025futurebench, qian2025agents}.
These benchmarks are not specific to crypto or financial advisory tasks, but are relevant due to shared characteristics such as uncertainty, time-varying information, and decision-making under incomplete knowledge.

These approaches are largely outcome-based, relying on extrinsic metrics such as forecasting accuracy or realized economic performance.
However, they are typically framed over general economic, political, and cross-asset settings rather than crypto-native or traditional-finance agent task suites.
Within that paradigm, forecasting benchmarks isolate predictive skill as information evolves over time~\citep{ye2024mirai, zeng2025futurex}.
Other benchmarks evaluate agents in prediction market venues or closely related simulated and live trading environments~\citep{arora2026predictionmarketbench, yu2025livetradebench, together2025futurebench, qian2025agents}.

While these benchmarks capture downstream performance in dynamic environments, they do not evaluate how agents communicate reasoning, structure outputs, or support user decision-making in human-in-the-loop settings.

\subsection{TradFi Agent Benchmarks}

Traditional finance (TradFi) agent benchmarks evaluate LLM agents on tasks such as financial question answering, analysis, and trading, using both reasoning-based and outcome-based criteria~\citep{islam2023financebench, xie2024finben, mateega2025financeqa, lu2025bizfinbench, li2025investorbench, qian2025agents, chen2025stockbench, fan2025ai, bigeard2025finance}.
These benchmarks are domain-specific to financial data and workflows, but are not tailored to crypto-native settings such as on-chain analysis or DeFi protocols.

Reasoning-based benchmarks stress financial knowledge and analysis on largely static, self-contained tasks (e.g., document-grounded question answering, structured finance problems) and score agent outputs with intrinsic criteria such as correctness and coherence~\citep{islam2023financebench, xie2024finben, mateega2025financeqa, lu2025bizfinbench}.
Outcome-based benchmarks evaluate multi-step or environment-coupled workflows such as trading or research pipelines, where success is often measured with extrinsic outcomes such as returns or end-to-end task completion~\citep{li2025investorbench, qian2025agents, chen2025stockbench, fan2025ai, bigeard2025finance}.

Although these benchmarks cover a wide range of financial tasks, they similarly do not operationalize how agents support human decision-making in advisory or copilot settings.

\subsection{Human-Centric Evaluation of LLMs and Agents}

A large methodological literature studies how to evaluate LLMs and agents using preference-based comparisons, structured rubrics, and LLM judges~\citep{zheng2023judging, chiang2024chatbot, li2023alpacaeval, chen2024t, tan2024judgebench, li2025generation, liu2023agentbench, du2025deepresearch, tang2025dsgbench, mohammadi2025evaluation, ferrag2025llm}.
This work focuses on evaluation methodology rather than domain-specific task design.

Preference-based frameworks compare model outputs using human or model judges, forming the basis of many modern leaderboards~\citep{zheng2023judging, chiang2024chatbot, li2023alpacaeval}.
Structured evaluation work studies rubric design, tool-use scoring, and judge reliability~\citep{chen2024t, tan2024judgebench}.
Broader agent benchmarks provide multi-task evaluation harnesses across environments~\citep{liu2023agentbench, du2025deepresearch}.

While this literature informs scalable evaluation pipelines, it does not define domain-specific criteria for decision support in crypto copilot settings. Our work builds on these methods to define structured, task-aware dimensions tailored to user-facing decision support.

\vspace{0.5em}

\begin{figure}[h]
\centering
\includegraphics[width=1.0\linewidth]{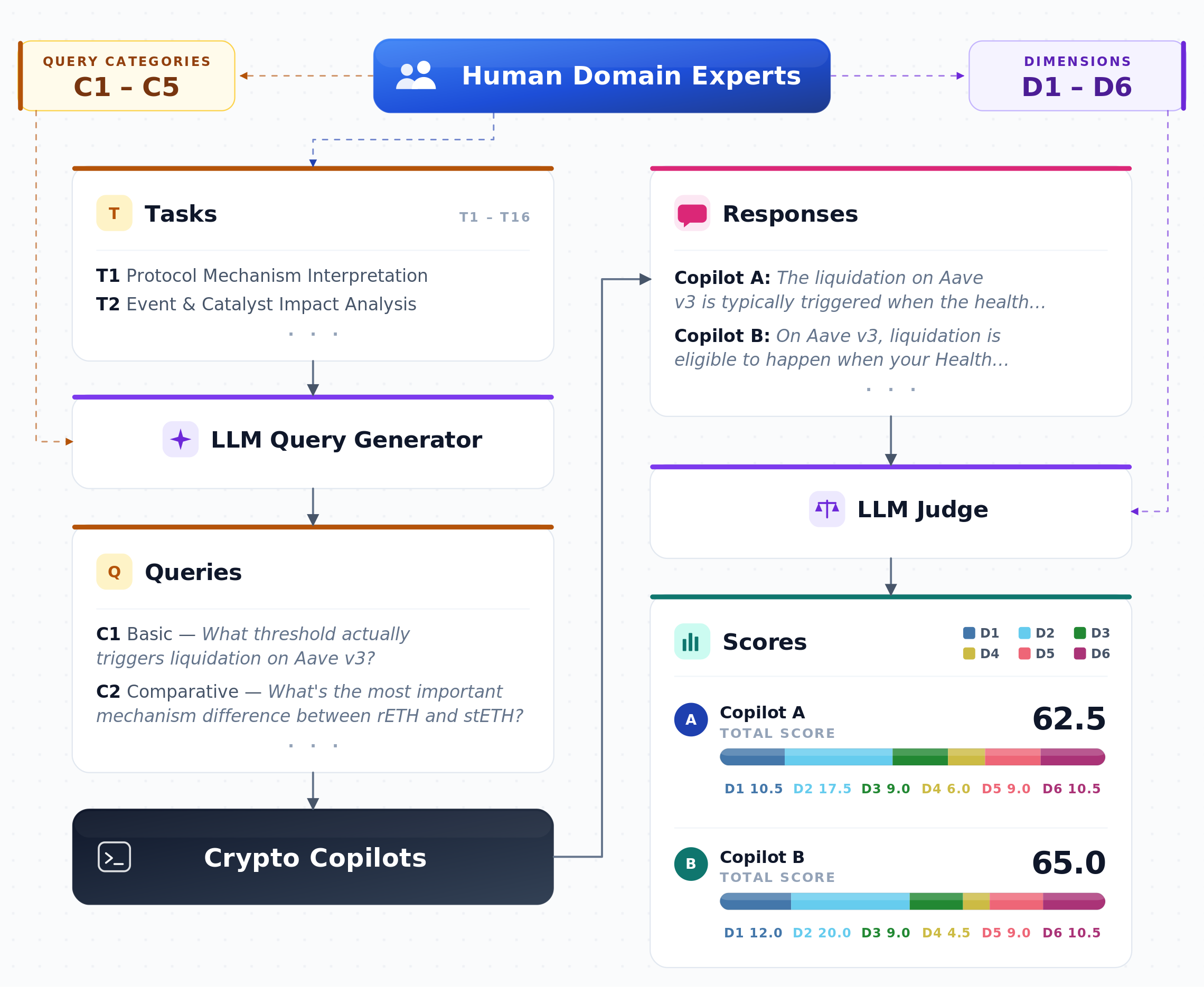}
\caption{
\textbf{Overview of \benchmark.}
The \benchmark\ benchmark is constructed through an iterative design process with input from domain experts, defining the evaluation dimensions, task taxonomy, query categories, LLM query generation prompts, and LLM judge rubrics.
For each of the 16 tasks, an LLM query generator produces multiple queries across five query categories.
These queries are executed on all tested crypto copilots, which produce final user-visible responses.
Each response is evaluated by an LLM judge using a shared rubric across six decision support dimensions.
Dimension-level scores are then aggregated using query-category-dependent weights to produce a final 0--100 score for each copilot response.
}
\label{fig:lattice}
\end{figure}

%% file: 3_benchmark.tex
\section{\benchmark}
\label{sec:benchmark}

\subsection{Overview}

\benchmark\ is designed to evaluate the decision support utility of crypto agents in realistic user-facing settings (Figure~\ref{fig:lattice}).
Rather than assessing agents' reasoning processes or realized outcomes, \benchmark\ focuses on whether agent outputs help users understand, evaluate, and act in human-in-the-loop workflows.

The evaluation inputs are intentionally lightweight: the judge observes only the user query, task/category metadata, and final (user-visible) agent response.
This design reflects how deployed crypto copilots are typically assessed in practice, and ensures that evaluation focuses on whether the response itself supports effective decision-making.

\benchmark\ is guided by three design principles:

\begin{itemize}
    \item \textbf{Decision support as a multi-dimensional property.}
    Rather than evaluating isolated correctness or outcomes, \benchmark\ decomposes decision support utility into interpretable dimensions that capture how users understand, evaluate, and act on crypto agent outputs.

    \item \textbf{Evaluation aligned with user-facing interactions.}
    Scoring is based on the user query, task context, and final response, reflecting how crypto copilots are used and assessed in practice.

    \item \textbf{Scalable, reusable evaluation.}
    The benchmark avoids per-example expert annotation and external data dependencies by using LLM judges and synthetic queries, enabling evaluation to scale across crypto agents and query sets without proportional increases in labeling effort.
\end{itemize}

\benchmark\ scores decision support utility along six evaluation dimensions, based on queries across 16 task types covering the end-to-end crypto copilot workflow and five query categories capturing core prompting patterns.
We describe each component below.

\subsection{Evaluation Dimensions}
\label{sec:benchmark:dimensions}

\benchmark\ evaluates each crypto agent response along six dimensions that capture complementary aspects of decision support utility (Table~\ref{tab:benchmark:dimensions}).
Together, these dimensions operationalize whether a response enables a user to understand the situation, assess available options, and take appropriate action in a human-in-the-loop setting.

The dimensions are designed to decompose decision support into interpretable components that can be evaluated independently.
Rather than treating response quality as a single scalar or relying on outcome-based metrics, this decomposition allows the benchmark to identify specific strengths and weaknesses in how an agent supports user decisions.
For example, a response may exhibit strong mechanism clarity but weak actionability, or provide useful guidance while failing to account for uncertainty.

\begin{table}[h]
\centering
\small
\renewcommand{\arraystretch}{1.2}
\begin{tabular}{clp{5.8cm}}
\toprule
\textbf{ID} & \textbf{Dimension} & \textbf{Description} \\
\midrule
D1 & Intent Fidelity & Aligns with user goals, constraints, and implicit intent without silently reframing the problem. \\ \addlinespace
D2 & Mechanism Clarity & Explains relevant mechanisms and causal relationships clearly and consistently. \\ \addlinespace
D3 & Uncertainty Handling & Represents uncertainty explicitly using scenarios, ranges, or conditional reasoning. \\ \addlinespace
D4 & Actionability & Provides concrete next steps, checks, or guardrails tied to the user’s objective. \\ \addlinespace
D5 & Evidence Coverage & Engages relevant evidence types and highlights important missing information. \\ \addlinespace
D6 & Response Structure & Organizes information clearly with consistent conclusions and no contradictions. \\
\bottomrule
\end{tabular}
\caption{
\textbf{\benchmark\ Evaluation Dimensions.}
Each dimension captures a distinct aspect of decision support utility, focusing on how effectively a crypto agent response supports user understanding, evaluation, and action.
}
\label{tab:benchmark:dimensions}
\end{table}

Notably, the dimensions are task-agnostic: they apply consistently across different types of crypto copilot interactions, from protocol analysis to execution planning.
At the same time, they are sufficiently granular to capture the different roles that responses play in supporting decisions, including aligning with user intent, structuring reasoning, surfacing uncertainty, and translating analysis into actionable guidance.

The dimensions are also designed to be assessable directly from the observable query--response pair, without requiring access to tool execution traces, live data, or external ground-truth verification.
This enables scalable evaluation using LLM judges, while ensuring that scores reflect the quality of the response as experienced by the user.

Each dimension is scored independently on a 0--10 scale by an LLM judge based on the user query, task context, and final response.
These scores are later aggregated using query-category-specific weights (\textsection~\ref{sec:benchmark:query_categories}) to produce a final score for each response.

\subsection{Evaluation Tasks}
\label{sec:benchmark:tasks}

\benchmark\ organizes queries into 16 task types that represent common user goals in crypto copilot interactions (Table~\ref{tab:benchmark:tasks}).
These tasks span the full lifecycle of decision-making, from understanding protocols and analyzing markets to assessing risks, planning execution, and updating beliefs based on feedback.

The task taxonomy defines the decision context in which a response is evaluated.
While evaluation dimensions capture how well a response supports decision-making, tasks specify what kind of decision or analysis the user is attempting to perform.
This separation allows the benchmark to evaluate responses in context, rather than as generic reasoning outputs.
For example, a high-quality response for mechanism interpretation differs substantially from one for execution planning, even if both are factually correct.

The tasks are designed to reflect realistic usage patterns rather than isolated capabilities.
Instead of evaluating narrow skills (e.g., entity extraction), \benchmark\ focuses on end-to-end user objectives that require combining reasoning, contextualization, and action-oriented guidance.
This ensures that evaluation captures how crypto agents function in practice as decision support tools.

Most tasks (T1--T15) correspond to standard analytical or decision-making scenarios encountered in crypto workflows.
Task T16 is treated separately and captures scope boundary recognition and safe refusal, where the appropriate behavior is to decline or redirect a request that is unsafe, out-of-domain, or otherwise inappropriate.
Including T16 ensures that the benchmark evaluates not only helpfulness, but also appropriate handling of unsafe or invalid requests.

Finally, the task set is designed to provide broad and balanced coverage of crypto copilot use cases.
The evaluation set includes all combinations of task types and query categories (\textsection~\ref{sec:benchmark:query_categories}), resulting in $16 \times 5 = 80$ task-category pairs.
This structured coverage enables systematic analysis of agent performance across both decision contexts and interaction patterns.

\begin{table}[t]
    \centering
    \small
    \renewcommand{\arraystretch}{1.2}
    \begin{tabular}{clp{5.8cm}}
    \toprule
    \textbf{ID} & \textbf{Task} & \textbf{Description} \\
    \midrule
    T1  & Protocol Mechanism Interpretation & Explain how a protocol or mechanism works and how it behaves under different conditions. \\ \addlinespace
    T2  & Event \& Catalyst Impact Analysis & Analyze how events or updates may affect markets, protocols, or positions. \\ \addlinespace
    T3  & Claim Validation \& Narrative Hygiene & Evaluate claims and distinguish verifiable facts from speculation or narrative. \\ \addlinespace
    T4  & Sentiment Structure \& Attention Diagnostics & Assess market sentiment and identify underlying drivers of attention or hype. \\ \addlinespace
    T5  & Market Regime \& Behavior Assessment & Characterize market conditions such as trend, volatility, or participation. \\ \addlinespace
    T6  & Liquidity, Capacity \& Execution Feasibility Analysis & Evaluate whether trades can be executed given liquidity, slippage, and market impact. \\ \addlinespace
    T7  & Derivative Pricing \& Positioning Interpretation & Interpret derivative signals such as funding or open interest to infer positioning. \\ \addlinespace
    T8  & Capital Flow \& Participation Analysis & Analyze how capital and activity move across markets or instruments. \\ \addlinespace
    T9  & Supply Dynamics \& Ownership Risk Assessment & Assess issuance, unlocks, and ownership concentration risks. \\ \addlinespace
    T10 & Risk Surface \& Failure Mode Identification & Identify key risks and potential failure modes relevant to a position or system. \\ \addlinespace
    T11 & Scenario-Based Stress \& Failure Analysis & Evaluate how systems or positions behave under adverse scenarios. \\ \addlinespace
    T12 & Participation \& Opportunity Cost Evaluation & Assess whether an opportunity is worth pursuing given risks and alternatives. \\ \addlinespace
    T13 & Capital Allocation \& Trade Structuring Under Constraints & Design allocations or trades under specified constraints. \\ \addlinespace
    T14 & Execution Planning \& Operational Safety Design & Plan execution steps with safeguards and validation checks. \\ \addlinespace
    T15 & Adaptive Learning from Market Feedback & Update strategies or beliefs based on observed outcomes or feedback. \\ \addlinespace
    T16 & Scope Boundary Recognition \& Safe Refusal & Identify out-of-scope or unsafe requests and respond appropriately. \\
    \bottomrule
    \end{tabular}
    \caption{
    \textbf{\benchmark\ Evaluation Tasks.}
    Each task represents a distinct type of user goal encountered in crypto copilot interactions, spanning analysis, decision-making, and execution.
    Descriptions summarize the core capability evaluated by each task.
    }
    \label{tab:benchmark:tasks}
    \end{table}

\subsection{Evaluation Query Categories}
\label{sec:benchmark:query_categories}

\benchmark\ defines five query categories that capture how a user frames a query (Table~\ref{tab:benchmark:categories}).
While tasks specify the underlying objective (i.e., \emph{what} the user is trying to do), query categories describe the interaction pattern (i.e., \emph{how} the user expresses that objective), such as whether a query is straightforward, comparative, or ambiguous.
This distinction allows the benchmark to evaluate how crypto agents adapt their responses to different types of user inputs within the same task.
For example, answering a constrained decision query requires different reasoning and structure than answering a basic explanatory query.

Importantly, different query types place different demands on decision support.
For instance, decision-oriented queries require strong uncertainty handling and actionable guidance, while comparative queries require clear structure and explicit trade-off analysis.
To reflect these differences, each query category defines a set of weights over the evaluation dimensions, specifying which aspects of a response are most important in that interaction context.

During evaluation, each dimension is first scored independently on a 0--10 scale.
These scores are then combined using the category-specific weights (Table~\ref{tab:benchmark:query_category_weights}) to produce a single 0--100 score for each query.
This weighted aggregation ensures that responses are evaluated not only on overall quality, but also on how well they prioritize the aspects of decision support that matter most for the given query type.

\begin{table}[t]
\centering
\small
\renewcommand{\arraystretch}{1.2}
\begin{tabular}{clp{5.8cm}}
\toprule
\textbf{ID} & \textbf{Query Category} & \textbf{Description} \\
\midrule
C1 & Basic & Direct factual or mechanism questions emphasizing clarity and explanation. \\ \addlinespace
C2 & Comparative & Queries that compare options and require explicit contrasts or structured trade-off analysis. \\ \addlinespace
C3 & Constrained & Queries with explicit constraints, such as capital limits, venue restrictions, or instrument preferences. \\ \addlinespace
C4 & Decision & Queries that ask for a recommendation or decision under uncertainty, requiring guidance and guardrails. \\ \addlinespace
C5 & Ambiguous & Under-specified queries that require clarification, cautious framing, or conservative assumptions. \\
\bottomrule
\end{tabular}
\caption{
\textbf{\benchmark\ Query Categories.}
Each category captures a different way that users frame requests within a given task. Categories are used both to diversify query types and to determine how evaluation dimensions are weighted during score aggregation.
}
\label{tab:benchmark:categories}
\end{table}

\begin{table}[t]
\centering
\small
\renewcommand{\arraystretch}{1.2}
\begin{tabular}{lcccccc}
\toprule
& \multicolumn{6}{c}{\textbf{Dimension}} \\
\cmidrule(lr){2-7}
\textbf{Query Category} & \textbf{D1} & \textbf{D2} & \textbf{D3} & \textbf{D4} & \textbf{D5} & \textbf{D6} \\
\midrule
C1 Basic       & 15 & 25 & 15 & 15 & 15 & 15 \\
C2 Comparative & 15 & 20 & 20 & 15 & 15 & 15 \\
C3 Constrained & 20 & 15 & 20 & 20 & 10 & 15 \\
C4 Decision    & 15 & 10 & 30 & 20 & 10 & 15 \\
C5 Ambiguous   & 25 & 15 & 25 & 10 & 10 & 15 \\
\bottomrule
\end{tabular}
\caption{
\textbf{\benchmark\ dimension weights by query category.}
Columns D1--D6 correspond to Intent Fidelity, Mechanism Clarity, Uncertainty Handling, Actionability, Evidence Coverage, and Response Structure, respectively.
Each row sums to 100 and defines how dimension scores are aggregated for a given query category.
}
\label{tab:benchmark:query_category_weights}
\end{table}

\subsection{Evaluation Protocol}

\paragraph{Inputs.}
We evaluate each crypto copilot on a fixed set of queries spanning all task types (\textsection~\ref{sec:benchmark:tasks}) and query categories (\textsection~\ref{sec:benchmark:query_categories}).
For each query, we collect the final user-visible response produced by each copilot, reflecting the output that a user would directly observe in a real interaction.

\paragraph{LLM Judge.}
Each response is evaluated by an LLM judge using a shared rubric based on the six evaluation dimensions (\textsection~\ref{sec:benchmark:dimensions}).
The judge receives the user query, task label, category label, and response text, and produces structured scores and rationales for each dimension.
These rationales provide interpretable explanations for each score, enabling qualitative analysis, debugging of agent behavior, and auditing of the evaluation process.
Crucially, scoring is performed independently for each copilot: the judge evaluates one response at a time without comparing it to others.
This avoids relative grading effects and ensures that scores reflect absolute decision support quality rather than stylistic differences between agents.
The use of structured dimension-level scoring provides more granular feedback than a single scalar score, enabling analysis of specific strengths and weaknesses in how agents support user decision-making.

\paragraph{Aggregation.}
For tasks T1--T15, dimension scores are first assigned independently on a 0--10 scale.
These scores are then combined using category-specific weights (Table~\ref{tab:benchmark:query_category_weights}) to produce a total score in the range 0--100.
This aggregation reflects the fact that different query types place different importance on aspects of decision support.
For example, decision-oriented queries place greater emphasis on uncertainty handling and actionability, while comparative queries emphasize structure and explicit trade-offs.
By incorporating category-dependent weights, the benchmark evaluates not only overall response quality, but also how well responses prioritize the aspects that matter most for the given interaction context.
For task T16 (safe refusal), only a single dimension (denoted as ``Refusal'') is evaluated, and the resulting score is scaled directly to the same 0--100 range.

\paragraph{Design Properties.}
The evaluation protocol is designed to be scalable, reproducible, and adaptable.
First, because scoring depends only on the observable query-response pair, the benchmark does not require per-example expert annotation or external ground-truth data.
This enables evaluation to scale across large query sets and multiple crypto agents without proportional increases in labeling effort.
Second, the use of a shared rubric and independent scoring ensures consistency across evaluations, allowing meaningful comparisons between agents and across different task and query types.
Finally, the rubric-based design allows the evaluation criteria to be iteratively refined over time.
As new interaction patterns, tasks, or evaluation needs emerge, the rubric can be updated without requiring reconstruction of the entire dataset, supporting continuous improvement of evaluation quality.

\paragraph{Benchmark Construction.}
The evaluation dimensions, task taxonomy, query categories, LLM query generation prompts, and LLM judge rubrics were developed through an iterative design process with input from three domain experts in crypto markets and trading workflows.
Across multiple rounds of refinement, we aligned the benchmark components with real-world crypto copilot usage patterns, focusing on how users interpret, assess, and act on crypto agent outputs.
This approach helps ensure that \benchmark\ captures domain-relevant notions of decision support, while enabling scalable evaluation through reusable rubrics rather than per-example expert labeling.
As an open-source benchmark, \benchmark\ can be continually extended and refined based on feedback from the broader community, allowing the evaluation framework to evolve alongside emerging use cases and standards.

%% file: 4_experiments.tex
\section{Experiments}
\label{sec:experiments}

\subsection{Overview}

Our main experiments follow the \benchmark\ protocol described in \textsection~\ref{sec:benchmark}.
We evaluate six crypto copilots: \href{https://www.heyelsa.ai/}{Elsa}, \href{https://www.askgina.ai/}{Gina}, \href{https://askjune.ai/}{June}, \href{https://minara.ai/}{Minara}, \href{https://heysorin.ai}{Sorin}, and \href{https://asksurf.ai/}{Surf}.
These copilots were selected as a mix of well-known and emerging products within the Web3 ecosystem, capturing a range of product designs and interaction styles for user-facing crypto decision support.

Our evaluation set consists of 1{,}200 queries spanning all task types and query categories, with 15 queries per task-category pair.
Consistent with the benchmark design, we treat each copilot as a black-box system and evaluate only its final user-visible responses.
All responses are scored using a fixed LLM judge and a shared rubric, producing structured scores across the six evaluation dimensions and aggregated totals for each query-copilot pair.
In these experiments, we use a GPT-5.2~\citep{openai_models_2025} model both to generate queries and as the LLM judge.

In the rest of this section, we report aggregate crypto copilot performance (\textsection~\ref{sec:experiments:agg_results}), followed by breakdowns with respect to evaluation dimensions (\textsection~\ref{sec:experiments:dim_results}), task types (\textsection~\ref{sec:experiments:task_results}), and query categories (\textsection~\ref{sec:experiments:cat_results}).
At a high level, the aggregate scores suggest a clear top performer and a tightly clustered mid-tier. 
However, this view masks meaningful differences in how copilots support decision-making.
Copilots with similar overall scores often diverge across dimensions such as uncertainty handling, actionability, and response structure, with these differences becoming more pronounced across different task types and query categories.

Furthermore, \textsection~\ref{sec:experiments:human_pref_study} presents a complementary human pairwise preferences study on a separate query set.
While \benchmark\ evaluates structured decision support properties using an LLM judge, the human study captures direct user preferences in head-to-head comparisons.
We use this study to assess whether the benchmark’s notion of decision support utility aligns with what users actually prefer in practice.
Although the two evaluations differ in setup and are not expected to match exactly, consistent patterns between them provide evidence that \benchmark\ captures meaningful aspects of real-world usefulness.

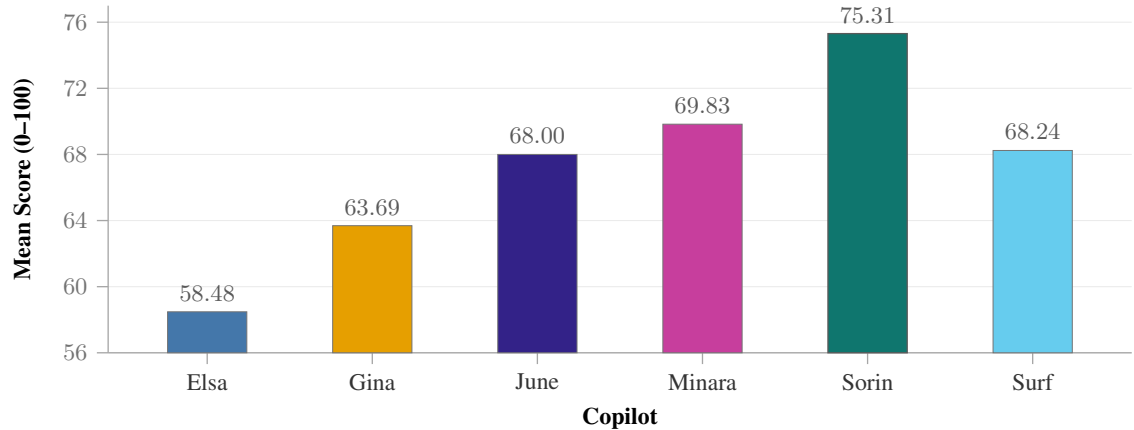
\begin{figure}[h]
	\centering
	\input{figures/chart_summary_bar}
	\caption{\textbf{Aggregate Results.} Per-copilot mean score (0--100).}
	\label{fig:summary_bar}
\end{figure}

\subsection{Aggregate Results}
\label{sec:experiments:agg_results}

Figure~\ref{fig:summary_bar} reports mean scores (0--100) over the 1{,}200 queries.
First, we find that Sorin achieves the highest aggregate score (75.31).
Next, we have a tightly clustered middle tier of Minara (69.83), Surf (68.24), and June (68.00)
Finally, Gina (63.69) and Elsa (58.48) form the lower tier.

The leaderboard exhibits a clear top-heavy structure: Sorin leads the next-best copilot by over five points, while differences within the middle tier are small (within 2 points).
This suggests that aggregate scores alone provide limited resolution among Minara, Surf, and June, despite meaningful differences in their underlying behavior.

At the same time, the separation between tiers indicates that the benchmark remains sensitive to larger differences in decision support quality.
In particular, Elsa consistently trails the rest of the copilots by a substantial margin, forming a distinct bottom band in the aggregate results.

Finally, aggregate scores partially reflect performance on boundary and refusal behavior (T16), which accounts for a non-trivial portion of the evaluation set.
However, as we show in later sections, these effects are more clearly understood through dimension- and task-level breakdowns than through the aggregate score alone.

\subsection{Dimension-Level Results}
\label{sec:experiments:dim_results}

Table~\ref{tab:experiments:dim_results} reports mean scores (0--10) for each evaluation dimension, defined in \textsection~\ref{sec:benchmark:dimensions}.
Dimensions D1--D6 are averaged over the 1{,}125 queries for tasks T1--T15, while Refusal is computed over the 75 queries for task T16.

Sorin leads across all standard dimensions (D1--D6), indicating consistently strong performance in core aspects of decision support.
However, the magnitude of this lead varies substantially by dimension.
Relative to Minara, June, and Surf, Sorin's margin is largest on response structure (D6), whereas uncertainty handling (D3) and actionability (D4) show tighter clustering among those three copilots.

At the same time, actionability (D4) spans a wide range over the full leaderboard, with larger gaps between copilots than in mechanism clarity (D2) or evidence coverage (D5).
This is driven largely by shortfalls for Elsa and Gina relative to the other four copilots.
Elsa, Gina, June, Sorin, and Surf all score higher on mechanism clarity than on actionability; Minara is nearly tied across the two (Table~\ref{tab:experiments:dim_results}), suggesting that competent explanations can coexist with comparatively weak guidance on concrete next steps.

In contrast, Refusal (scope boundary recognition and safe refusal) exhibits a different pattern: June and Surf lead this dimension, while Sorin and Gina trail despite strong performance elsewhere.
This indicates that boundary handling represents a distinct capability that does not necessarily correlate with overall decision support performance.

Overall, the dimension-level results reinforce that decision support utility is inherently multi-faceted.
Copilots that appear similar in aggregate scores---such as Minara, Surf, and June---can differ meaningfully in how they balance explanation, uncertainty, and actionability, underscoring the importance of evaluating these dimensions separately.

\begin{table}[t]
\centering
\small
\setlength{\tabcolsep}{4pt}
\renewcommand{\arraystretch}{1.15}
\begin{tabular}{lccccccc}
\toprule
& \multicolumn{7}{c}{\textbf{Dimension}} \\
\cmidrule(lr){2-8}
\textbf{Copilot} & \textbf{D1} & \textbf{D2} & \textbf{D3} & \textbf{D4} & \textbf{D5} & \textbf{D6} & \textbf{Refusal} \\
\midrule
Elsa   & 6.78 & 5.90 & 5.11 & 4.69 & 5.02 & 6.43 & 8.52 \\
Gina   & 6.55 & 6.74 & 5.83 & 5.80 & 5.96 & 7.26 & 6.65 \\
June   & 7.17 & 6.94 & 6.10 & 6.18 & 6.14 & 7.43 & \textbf{8.91} \\
Minara & 7.23 & 6.90 & 6.90 & 6.88 & 6.71 & 6.92 & 7.61 \\
Sorin  & \textbf{7.99} & \textbf{7.88} & \textbf{6.93} & \textbf{7.04} & \textbf{6.88} & \textbf{8.78} & 6.67 \\
Surf   & 6.59 & 6.50 & 6.39 & 6.38 & 6.86 & 7.74 & 8.43 \\
\bottomrule
\end{tabular}
\caption{
\textbf{Dimension-Level Results.}
Average dimension score (0--10) per copilot.
Columns D1--D6 correspond to Intent Fidelity, Mechanism Clarity, Uncertainty Handling, Actionability, Evidence Coverage, and Response Structure.
Refusal reports the score for the safe refusal dimension.
}
\label{tab:experiments:dim_results}
\end{table}

\subsection{Task-Level Results}
\label{sec:experiments:task_results}

Table~\ref{tab:experiments:task_results} reports mean scores (0--100) for each copilot across the sixteen task types (T1--T16), as described in \textsection~\ref{sec:benchmark:tasks}.
Sorin achieves the highest mean score across all standard tasks (T1--T15), while June leads T16, followed by Surf and Elsa.
Compared to aggregate results, the task-level view provides a more granular perspective on how copilots perform across different decision contexts.

\begin{table}[t]
	\centering
	\small
	\begingroup
	\setlength{\tabcolsep}{4.5pt}%
	\begin{tabular*}{\linewidth}{@{\extracolsep{\fill}}l*{16}{c}@{}}
		\toprule
		\multirow{2}{*}{Copilot} & \multicolumn{16}{c}{Task} \\
		\cmidrule(lr){2-17}
		& T1 & T2 & T3 & T4 & T5 & T6 & T7 & T8 & T9 & T10 & T11 & T12 & T13 & T14 & T15 & T16 \\
		\midrule
		Elsa   & 54.5 & 55.4 & 59.1 & 58.2 & 56.6 & 57.5 & 56.9 & 57.6 & 54.5 & 57.8 & 56.6 & 56.9 & 56.5 & 57.2 & 55.2 & 85.2 \\
		Gina   & 65.0 & 62.0 & 61.0 & 55.6 & 61.8 & 61.1 & 60.3 & 65.2 & 64.0 & 67.3 & 63.7 & 67.2 & 63.5 & 68.8 & 66.1 & 66.5 \\
		June   & 63.8 & 64.8 & 66.3 & 64.8 & 67.0 & 67.6 & 69.6 & 65.6 & 64.5 & 64.9 & 65.2 & 65.7 & 69.8 & 68.9 & 70.3 & \textbf{89.1} \\
		Minara & 66.1 & 68.5 & 73.3 & 68.2 & 68.6 & 71.1 & 70.1 & 69.9 & 71.3 & 69.1 & 70.7 & 68.2 & 68.1 & 69.5 & 68.6 & 76.1 \\
		Sorin  & \textbf{75.3} & \textbf{75.3} & \textbf{76.5} & \textbf{76.7} & \textbf{75.5} & \textbf{75.6} & \textbf{76.3} & \textbf{75.9} & \textbf{75.6} & \textbf{77.0} & \textbf{75.9} & \textbf{75.7} & \textbf{74.5} & \textbf{75.7} & \textbf{76.8} & 66.7 \\
		Surf   & 68.5 & 67.4 & 67.5 & 68.7 & 69.0 & 65.3 & 67.4 & 67.8 & 67.1 & 67.4 & 67.7 & 66.1 & 66.7 & 65.2 & 65.7 & 84.3 \\
		\bottomrule
	\end{tabular*}
	\endgroup
	\caption{\textbf{Task-Level Results.} Average score (0--100) per copilot per task (mean over seventy-five queries per task: fifteen per task--category cell). Task names abbreviated; see Table~\ref{tab:benchmark:tasks} for full names.}
	\label{tab:experiments:task_results}
\end{table}

The task-level results reveal that performance depends strongly on the decision context.
While Sorin consistently leads across analytical and execution-oriented tasks (T1--T15), no single copilot dominates across all task types.
In particular, T16 (scope boundary recognition and safe refusal) exhibits a distinct pattern: June and Surf achieve the highest scores, while Sorin and Gina rank lower despite strong performance elsewhere.

This divergence indicates that boundary handling constitutes a separate capability that does not necessarily correlate with overall decision support performance.
More broadly, variation across tasks suggests that copilots may be optimized for different parts of the decision-making workflow, such as analysis, execution planning, or safety-related interactions.

Finally, the task-level view highlights that differences between copilots are often localized rather than uniform.
For example, Elsa scores near the bottom on most T1--T15 tasks yet posts a comparatively high T16 score (85.2), while Gina's performance swings by more than ten points across standard tasks (e.g., 55.6 on T4 versus 68.8 on T14).
These kinds of heterogeneous patterns are not visible in aggregate scores, reinforcing the importance of evaluating copilots across diverse task types.

\subsection{Query-Category-Level Results}
\label{sec:experiments:cat_results}

\begin{figure}[t]
	\centering
	\input{figures/chart_category_bar}
	\captionsetup{skip=5pt}
	\caption{\textbf{Query-Category-Level Results.} Mean score (0--100) by query category.}
	\label{fig:experiments:cat_results}
\end{figure}
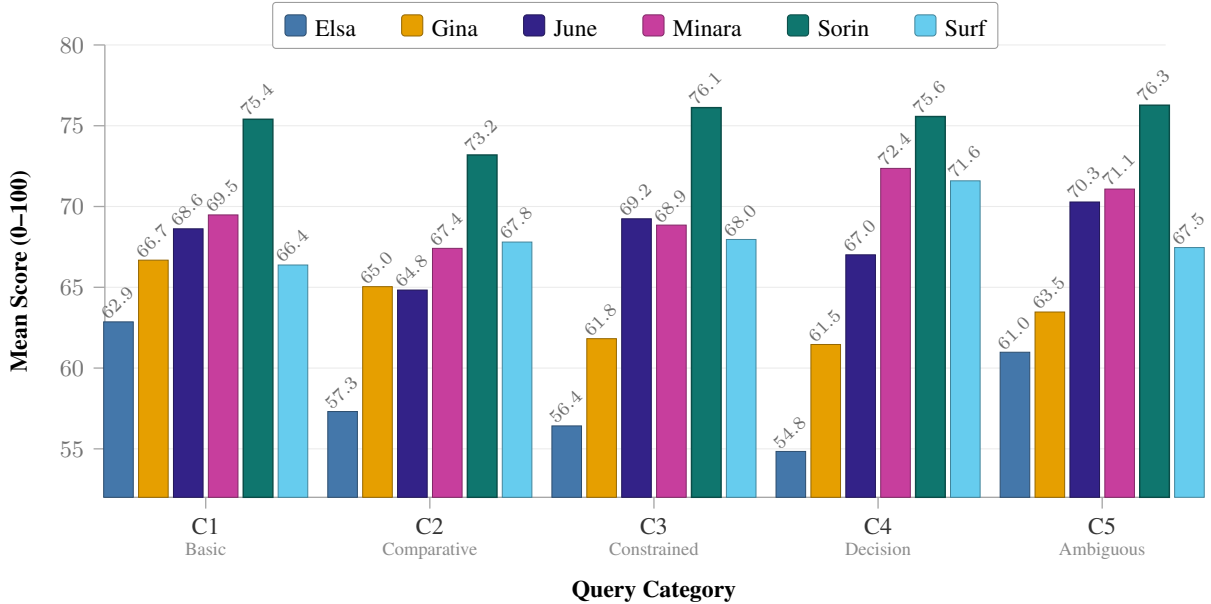

Figure~\ref{fig:experiments:cat_results} reports mean total scores (0--100) for each copilot across the five query categories (\textsection~\ref{sec:benchmark:query_categories}).
Sorin achieves the highest score in every category, while Minara is the runner-up in Basic (C1), Decision (C4), and Ambiguous (C5).
June is the runner-up in Constrained (C3), whereas Surf ranks second in Comparative (C2).

Performance varies systematically across categories, consistent with the category-dependent weighting scheme given in Table~\ref{tab:benchmark:query_category_weights}.
In particular, constrained queries produce the widest separation across copilots, with Sorin reaching 76.12 and Elsa trailing at 56.42, suggesting that explicit requirements and trade-offs remain a strong differentiator.
By contrast, comparative and decision-oriented queries produce a more compressed mid-tier, with Minara and Surf especially close in C4 (72.36 vs.\ 71.59).

The category-level results also reveal that copilot behavior is not uniform across categories.
For example, Surf performs strongest on C4 and weakest on C1, indicating relatively stronger support for recommendation-style queries than for straightforward explanatory ones.
Similarly, June performs best on C5 and weakest on C2, suggesting relative strength in under-specified settings that require clarification or cautious framing, compared with prompts that emphasize head-to-head comparison.

Overall, these results reinforce that decision support utility depends not only on the underlying task, but also on how the user frames the request.
Aggregate scores alone obscure these interaction-specific differences, which become visible only when conditioning on query category.

\subsection{Human Preferences Study}
\label{sec:experiments:human_pref_study}

We conduct a separate human pairwise preference study to complement \benchmark's LLM judge evaluation results (\textsection~\ref{sec:experiments:agg_results}--\textsection~\ref{sec:experiments:cat_results}).
Annotators are shown blinded pairs of responses to 80 crypto queries---spanning the same task types and query categories used by \benchmark---and asked to select a winner (or tie), along with a single main reason for their choice.
Across all copilot pairs, we collect 18{,}481 judgments, corresponding to roughly eleven annotators per query-copilot pair.

We define win rate as the fraction of pairwise comparisons in which a copilot's response is preferred (excluding ties, which account for 8.3\% of judgments).
Table~\ref{tab:experiments:human_pref_study:win} reports win rates across the full set of copilots.
At a high level, the human results are broadly consistent with \benchmark's LLM judge evaluation at the top of the leaderboard: Sorin achieves the highest win rate, aligning with its leading aggregate score.
However, differences emerge within the mid-tier.
Surf ranks second by human preference, while Minara (ranked second under \benchmark\ rubric) falls to fourth in the crowd evaluation.
Similarly, Gina performs relatively better with human annotators than under the rubric.

\begin{table}[t]
	\centering
	\begin{tabular}{lcc}
		\toprule
		\textbf{Copilot} & \textbf{Win Rate (\%)} & \textbf{Wins / Appearances} \\
		\midrule
		Elsa     & 18.03 & 952 / 5280 \\
		Gina     & 54.34 & 2869 / 5280 \\
		June     & 36.42 & 1923 / 5280 \\
		Minara   & 43.72 & 2309 / 5281 \\
		Sorin    & \textbf{62.68} & \textbf{3310 / 5281} \\
		Surf     & 55.55 & 2933 / 5280 \\
		Wayfinder & 50.19 & 2650 / 5280 \\
		\bottomrule
	\end{tabular}
	\caption{\textbf{Human Pairwise Win Rates.} Results over 18{,}481 judgments across all copilot pairs.}
	\label{tab:experiments:human_pref_study:win}
\end{table}

These discrepancies suggest that while the \benchmark\ captures structured aspects of decision support, human preferences place additional emphasis on perceived usefulness and interaction quality.
This interpretation is supported by the distribution of annotator-selected reasons (Table~\ref{tab:experiments:human_pref_study:main_reason}), where usefulness accounts for the largest share of preferences (45.54\%), followed by reasoning (35.37\%) and accuracy (19.08\%).
Despite these differences, the overall ordering remains partially aligned, particularly at the top and bottom of the leaderboard.
This suggests that \benchmark\ captures meaningful aspects of real-world usefulness, while also revealing differences in how structured evaluation and human preference weight various aspects of decision support.

\begin{table}[t]
	\centering
	\small
	\begin{tabular}{p{0.52\linewidth}rr}
		\toprule
		\textbf{Main Reason} & \textbf{Share (\%)} & \textbf{Count} \\
		\midrule
		Accuracy -- more correct / fewer factual errors & 19.08 & 3234 \\
		Usefulness -- more actionable and directly answers the user's need & \textbf{45.54} & \textbf{7718} \\
		Reasoning -- better logic, tradeoffs, and justification & 35.37 & 5994 \\
		Caution -- better handling of risk, uncertainty, and assumptions & 0.00 & 0 \\
		Clarity -- easier to read, better structured, less confusing & 0.00 & 0 \\
		\bottomrule
	\end{tabular}
	\caption{\textbf{Main Reasons for Human Preferences.} Distribution of annotator-selected reasons for preferring the winning response (non-tie judgments).}
	\label{tab:experiments:human_pref_study:main_reason}
\end{table}

Finally, we analyze serious errors in rejected responses.
Table~\ref{tab:human_serious} reports the fraction of losses in which annotators identified a major issue.
Across copilots, serious errors are flagged in roughly half to two-thirds of losses, with missed user intent, misleading presentation, and factual errors accounting for the majority of cases.

\begin{table}[t]
	\centering
	\begin{tabular}{lcc}
		\toprule
		\textbf{Copilot} & \textbf{Serious Error Rate When Lost (\%)} & \textbf{Flags / Losses} \\
		\midrule
		Elsa      & 58.90 & 2340 / 3973 \\
		Gina      & 58.35 & 1115 / 1911 \\
		June      & 50.80 & 1496 / 2945 \\
		Minara    & 62.86 & 1606 / 2555 \\
		Sorin     & 48.99 & 728 / 1486 \\
		Surf      & 52.06 & 974 / 1871 \\
		Wayfinder & \textbf{48.75} & 1075 / 2205 \\
		\bottomrule
	\end{tabular}
	\caption{\textbf{Serious Error Rates on Rejected Responses.} Fraction of losses in which annotators identified a major issue, computed over non-tie judgments.}
	\label{tab:human_serious}
\end{table}

Overall, the human study provides a complementary perspective on copilot performance.
While not directly comparable to \benchmark's LLM judge scores due to differences in queries and evaluation setup, the results offer evidence that the benchmark’s notion of decision support utility aligns with key aspects of user preference, while also highlighting areas where structured evaluation and human judgment diverge.

%% file: figures/chart_summary_bar.tex
\begin{tikzpicture}
\begin{axis}[
	experiment bar axis,
	width=0.92\linewidth,
	height=6.2cm,
	ymin=56,
	ymax=77,
	ylabel={Mean Score (0--100)},
	ylabel style={font=\small\bfseries, yshift=2pt},
	ytick={56,60,64,68,72,76},
	yticklabel style={
		font=\small,
		black!55,
		/pgf/number format/fixed,
		/pgf/number format/precision=0,
		/pgf/number format/fixed zerofill=false,
	},
	enlarge x limits={0.12},
	xtick={1,2,3,4,5,6},
	xticklabels={Elsa,Gina,June,Minara,Sorin,Surf},
	xticklabel style={font=\small, black!80},
	xlabel={Copilot},
	xlabel style={font=\small\bfseries},
]
\addplot[draw=none, forget plot] coordinates {(1,56) (6,77)};
\pgfplotsextra{
	\pgfmathsetmacro{\SummaryYmin}{56}%
	\pgfmathsetmacro{\BarHalfW}{0.24}%
	\fill[fill=cpElsa, draw=black!55, line width=0.35pt]
		(axis cs:{1-\BarHalfW},\SummaryYmin) rectangle (axis cs:{1+\BarHalfW},58.48);
	\fill[fill=cpGina, draw=black!55, line width=0.35pt]
		(axis cs:{2-\BarHalfW},\SummaryYmin) rectangle (axis cs:{2+\BarHalfW},63.69);
	\fill[fill=cpJune, draw=black!55, line width=0.35pt]
		(axis cs:{3-\BarHalfW},\SummaryYmin) rectangle (axis cs:{3+\BarHalfW},68.00);
	\fill[fill=cpMinara, draw=black!55, line width=0.35pt]
		(axis cs:{4-\BarHalfW},\SummaryYmin) rectangle (axis cs:{4+\BarHalfW},69.83);
	\fill[fill=cpSorin, draw=black!65, line width=0.5pt]
		(axis cs:{5-\BarHalfW},\SummaryYmin) rectangle (axis cs:{5+\BarHalfW},75.31);
	\fill[fill=cpSurf, draw=black!55, line width=0.35pt]
		(axis cs:{6-\BarHalfW},\SummaryYmin) rectangle (axis cs:{6+\BarHalfW},68.24);
	\node[font=\small, text=black!65, anchor=south, inner sep=2pt, yshift=2pt]
		at (axis cs:1,58.48) {\pgfmathprintnumber[fixed,precision=2,fixed zerofill=true]{58.48}};
	\node[font=\small, text=black!65, anchor=south, inner sep=2pt, yshift=2pt]
		at (axis cs:2,63.69) {\pgfmathprintnumber[fixed,precision=2,fixed zerofill=true]{63.69}};
	\node[font=\small, text=black!65, anchor=south, inner sep=2pt, yshift=2pt]
		at (axis cs:3,68.00) {\pgfmathprintnumber[fixed,precision=2,fixed zerofill=true]{68.00}};
	\node[font=\small, text=black!65, anchor=south, inner sep=2pt, yshift=2pt]
		at (axis cs:4,69.83) {\pgfmathprintnumber[fixed,precision=2,fixed zerofill=true]{69.83}};
	\node[font=\small, text=black!65, anchor=south, inner sep=2pt, yshift=2pt]
		at (axis cs:5,75.31) {\pgfmathprintnumber[fixed,precision=2,fixed zerofill=true]{75.31}};
	\node[font=\small, text=black!65, anchor=south, inner sep=2pt, yshift=2pt]
		at (axis cs:6,68.24) {\pgfmathprintnumber[fixed,precision=2,fixed zerofill=true]{68.24}};
}
\end{axis}
\end{tikzpicture}

%% file: figures/chart_category_bar.tex
\begin{tikzpicture}
\begin{axis}[
	name=catbar,
	experiment bar axis,
	width=0.95\linewidth,
	height=7.6cm,
	ybar,
	ymin=52,
	ymax=80,
	ylabel={Mean Score (0--100)},
	ylabel style={font=\small\bfseries, yshift=2pt},
	enlarge x limits=false,
	xmin=-0.35,
	xmax=13.85,
	bar width=11.2pt,
	xtick={1,4,7,10,13},
	xticklabels={
		\shortstack{C1\\\scriptsize\color{black!45}Basic},
		\shortstack{C2\\\scriptsize\color{black!45}Comparative},
		\shortstack{C3\\\scriptsize\color{black!45}Constrained},
		\shortstack{C4\\\scriptsize\color{black!45}Decision},
		\shortstack{C5\\\scriptsize\color{black!45}Ambiguous},
	},
	xticklabel style={align=center},
	xlabel={Query Category},
	xlabel style={
		font=\small\bfseries,
		at={(rel axis cs:0.5176,0)},
		anchor=north,
		yshift=-28pt,
	},
	nodes near coords,
	nodes near coords style={
		font=\scriptsize,
		text=black!58,
		inner sep=1pt,
		rotate=45,
		anchor=south west,
		xshift=-3.5pt,
		yshift=2pt,
		/pgf/number format/fixed,
		/pgf/number format/precision=1,
		/pgf/number format/fixed zerofill=true,
	},
	legend style={
		overlay,
		opacity=0,
		draw=none,
		fill=none,
	},
]
\addplot[fill=cpElsa,   draw=cpElsa!65!black,   line width=0.35pt] coordinates {(1,62.86) (4,57.31) (7,56.42) (10,54.84) (13,60.98)};
\addlegendentry{Elsa}
\addplot[fill=cpGina,   draw=cpGina!65!black,   line width=0.35pt] coordinates {(1,66.68) (4,65.04) (7,61.82) (10,61.46) (13,63.47)};
\addlegendentry{Gina}
\addplot[fill=cpJune,   draw=cpJune!65!black,   line width=0.35pt] coordinates {(1,68.62) (4,64.83) (7,69.24) (10,67.01) (13,70.28)};
\addlegendentry{June}
\addplot[fill=cpMinara, draw=cpMinara!65!black, line width=0.35pt] coordinates {(1,69.48) (4,67.41) (7,68.85) (10,72.36) (13,71.08)};
\addlegendentry{Minara}
\addplot[fill=cpSorin,  draw=cpSorin!65!black,  line width=0.55pt] coordinates {(1,75.40) (4,73.19) (7,76.12) (10,75.57) (13,76.27)};
\addlegendentry{Sorin}
\addplot[fill=cpSurf,   draw=cpSurf!65!black,   line width=0.35pt] coordinates {(1,66.38) (4,67.80) (7,67.96) (10,71.59) (13,67.46)};
\addlegendentry{Surf}
\end{axis}
\node[
	anchor=south,
	font=\footnotesize,
	draw=black!40,
	line width=0.4pt,
	fill=white,
	rounded corners=1.25pt,
	inner sep=4.5pt,
	yshift=-1.5pt,
] at (catbar.north) {%
	\tikz[baseline=0pt]{\draw[fill=cpElsa, draw=cpElsa!65!black, line width=0.28pt] (0,0) rectangle (8pt,8pt);}%
	\kern3.25pt Elsa\hspace{1.9em}%
	\tikz[baseline=0pt]{\draw[fill=cpGina, draw=cpGina!65!black, line width=0.28pt] (0,0) rectangle (8pt,8pt);}%
	\kern3.25pt Gina\hspace{1.9em}%
	\tikz[baseline=0pt]{\draw[fill=cpJune, draw=cpJune!65!black, line width=0.28pt] (0,0) rectangle (8pt,8pt);}%
	\kern3.25pt June\hspace{1.9em}%
	\tikz[baseline=0pt]{\draw[fill=cpMinara, draw=cpMinara!65!black, line width=0.28pt] (0,0) rectangle (8pt,8pt);}%
	\kern3.25pt Minara\hspace{1.9em}%
	\tikz[baseline=0pt]{\draw[fill=cpSorin, draw=cpSorin!65!black, line width=0.4pt] (0,0) rectangle (8pt,8pt);}%
	\kern3.25pt Sorin\hspace{1.9em}%
	\tikz[baseline=0pt]{\draw[fill=cpSurf, draw=cpSurf!65!black, line width=0.28pt] (0,0) rectangle (8pt,8pt);}%
	\kern3.25pt Surf%
};
\end{tikzpicture}

%% file: 5_conclusion.tex
\section{Conclusion}
\label{sec:conclusion}

We introduced \benchmark, a benchmark for evaluating the decision support utility of crypto agents in realistic, user-facing copilot settings.
Across six deployed copilots and 1{,}200 queries, our results show that aggregate scores alone provide an incomplete picture of performance.
While a clear top performer emerges, several copilots cluster closely in the mid-tier, despite exhibiting distinct strengths and weaknesses across evaluation dimensions, task types, and query categories.

These findings highlight that decision support utility is inherently context-dependent.
Copilots that perform similarly in aggregate can differ substantially in how they handle uncertainty, structure responses, or provide actionable guidance, with these differences becoming more pronounced under specific tasks and interaction patterns.
In particular, performance varies systematically with both the underlying task and how the query is framed, reinforcing the importance of evaluating across diverse decision contexts rather than relying on a single summary metric.

The human preference study provides a complementary perspective on these results.
While broadly consistent with the LLM-judge evaluation at the top of the leaderboard, it reveals meaningful reordering within the mid-tier, suggesting that structured evaluation and human judgment emphasize different aspects of response quality.
Annotators most frequently cite usefulness and reasoning as the primary drivers of preference, underscoring the importance of practical relevance and interpretability in real-world decision support.

To support reproducible research, we open-source all \benchmark\ code and data used in this paper.

\textbf{Limitations.}
Our results depend on the choice of LLM judge, rubric, and prompting setup, so alternative configurations may shift absolute scores or relative margins.
The evaluation is based on static query-response pairs and does not incorporate live data access, execution outcomes, or factual verification beyond what is observable in the response.
In addition, each query-copilot pair is evaluated once, without accounting for response variability.

\textbf{Future Work.}
Future work includes analyzing judge--human agreement at a finer granularity across tasks, categories, and preference rationales, as well as extending the benchmark to additional copilots and broader query sets.
Expanding the evaluation framework itself---including the set of dimensions, task taxonomy, and query categories---would further improve coverage of real-world decision support scenarios.
From a systems perspective, building a streamlined evaluation pipeline would make it easier for researchers and developers to use to test crypto agents using \benchmark, enabling faster iteration and more standardized comparisons.
Methodologically, incorporating judge ensembles and calibration techniques could improve robustness while preserving scalability.
Beyond static chat transcripts, repeated sampling, response-variability analysis, and selective verification against external references would better approximate copilot behavior in settings where answers depend on live tools or changing market data.

%% file: references.bib
@article{dai2025hallucination,
  title={When Hallucination Costs Millions: Benchmarking AI Agents in High-Stakes Adversarial Financial Markets},
  author={Dai, Zeshi and Peng, Zimo and Cheng, Zerui and Li, Ryan Yihe},
  journal={arXiv preprint arXiv:2510.00332},
  year={2025}
}

@article{guo2025cryptobench,
  title={CryptoBench: A Dynamic Benchmark for Expert-Level Evaluation of LLM Agents in Cryptocurrency},
  author={Guo, Jiacheng and Huang, Suozhi and Yao, Zixin and Zhang, Yifan and Lu, Yifu and Liu, Jiashuo and Li, Zihao and Deng, Nicholas and Xiao, Qixin and Tian, Jia and others},
  journal={arXiv preprint arXiv:2512.00417},
  year={2025}
}

@article{xie2024finben,
  title={Finben: A holistic financial benchmark for large language models},
  author={Xie, Qianqian and Han, Weiguang and Chen, Zhengyu and Xiang, Ruoyu and Zhang, Xiao and He, Yueru and Xiao, Mengxi and Li, Dong and Dai, Yongfu and Feng, Duanyu and others},
  journal={Advances in Neural Information Processing Systems},
  volume={37},
  pages={95716--95743},
  year={2024}
}

@article{du2025deepresearch,
  title={Deepresearch bench: A comprehensive benchmark for deep research agents},
  author={Du, Mingxuan and Xu, Benfeng and Zhu, Chiwei and Wang, Xiaorui and Mao, Zhendong},
  journal={arXiv preprint arXiv:2506.11763},
  year={2025}
}

@inproceedings{li2025investorbench,
  title={Investorbench: A benchmark for financial decision-making tasks with llm-based agent},
  author={Li, Haohang and Cao, Yupeng and Yu, Yangyang and Javaji, Shashidhar Reddy and Deng, Zhiyang and He, Yueru and Jiang, Yuechen and Zhu, Zining and Subbalakshmi, Kp and Huang, Jimin and others},
  booktitle={Proceedings of the 63rd Annual Meeting of the Association for Computational Linguistics (Volume 1: Long Papers)},
  pages={2509--2525},
  year={2025}
}

@article{qian2025agents,
  title={When agents trade: Live multi-market trading benchmark for llm agents},
  author={Qian, Lingfei and Peng, Xueqing and Wang, Yan and Zhang, Vincent Jim and He, Huan and Smith, Hanley and Han, Yi and He, Yueru and Li, Haohang and Cao, Yupeng and others},
  journal={arXiv preprint arXiv:2510.11695},
  year={2025}
}

@article{bigeard2025finance,
  title={Finance agent benchmark: Benchmarking llms on real-world financial research tasks},
  author={Bigeard, Antoine and Nashold, Langston and Krishnan, Rayan and Wu, Shirley},
  journal={arXiv preprint arXiv:2508.00828},
  year={2025}
}

@article{fan2025ai,
  title={AI-Trader: Benchmarking Autonomous Agents in Real-Time Financial Markets},
  author={Fan, Tianyu and Yang, Yuhao and Jiang, Yangqin and Zhang, Yifei and Chen, Yuxuan and Huang, Chao},
  journal={arXiv preprint arXiv:2512.10971},
  year={2025}
}

@article{chen2025stockbench,
  title={Stockbench: Can llm agents trade stocks profitably in real-world markets?},
  author={Chen, Yanxu and Yao, Zijun and Liu, Yantao and Ye, Jin and Yu, Jianing and Hou, Lei and Li, Juanzi},
  journal={arXiv preprint arXiv:2510.02209},
  year={2025}
}

@article{tang2025dsgbench,
  title={Dsgbench: A diverse strategic game benchmark for evaluating llm-based agents in complex decision-making environments},
  author={Tang, Wenjie and Zhou, Yuan and Xu, Erqiang and Cheng, Keyan and Li, Minne and Xiao, Liquan},
  journal={arXiv preprint arXiv:2503.06047},
  year={2025}
}

@inproceedings{mohammadi2025evaluation,
  title={Evaluation and benchmarking of llm agents: A survey},
  author={Mohammadi, Mahmoud and Li, Yipeng and Lo, Jane and Yip, Wendy},
  booktitle={Proceedings of the 31st ACM SIGKDD Conference on Knowledge Discovery and Data Mining V. 2},
  pages={6129--6139},
  year={2025}
}

@article{ferrag2025llm,
  title={From llm reasoning to autonomous ai agents: A comprehensive review},
  author={Ferrag, Mohamed Amine and Tihanyi, Norbert and Debbah, Merouane},
  journal={arXiv preprint arXiv:2504.19678},
  year={2025}
}

@article{tan2024judgebench,
  title={Judgebench: A benchmark for evaluating llm-based judges},
  author={Tan, Sijun and Zhuang, Siyuan and Montgomery, Kyle and Tang, William Y and Cuadron, Alejandro and Wang, Chenguang and Popa, Raluca Ada and Stoica, Ion},
  journal={arXiv preprint arXiv:2410.12784},
  year={2024}
}

@article{luo2025llm,
  title={Llm-powered multi-agent system for automated crypto portfolio management},
  author={Luo, Yichen and Feng, Yebo and Xu, Jiahua and Tasca, Paolo and Liu, Yang},
  journal={arXiv preprint arXiv:2501.00826},
  year={2025}
}

@article{liu2023agentbench,
  title={Agentbench: Evaluating llms as agents},
  author={Liu, Xiao and Yu, Hao and Zhang, Hanchen and Xu, Yifan and Lei, Xuanyu and Lai, Hanyu and Gu, Yu and Ding, Hangliang and Men, Kaiwen and Yang, Kejuan and others},
  journal={arXiv preprint arXiv:2308.03688},
  year={2023}
}

@article{islam2023financebench,
  title={FinanceBench: A new benchmark for financial question answering. arXiv},
  author={Islam, Pranab and Kannappan, Anand and Kiela, Douwe and Qian, Rebecca and Scherrer, Nino and Vidgen, Bertie},
  journal={arXiv preprint arXiv:2311.11944},
  year={2023}
}

@article{eswaran2026cryptoanalystbench,
  title={CryptoAnalystBench: Failures in Multi-Tool Long-Form LLM Analysis},
  author={Eswaran, Anushri and Golev, Oleg and Tank, Darshan and Rahi, Sidhant and Tyagi, Himanshu},
  journal={arXiv preprint arXiv:2602.11304},
  year={2026}
}

@inproceedings{li2025generation,
  title={From generation to judgment: Opportunities and challenges of llm-as-a-judge},
  author={Li, Dawei and Jiang, Bohan and Huang, Liangjie and Beigi, Alimohammad and Zhao, Chengshuai and Tan, Zhen and Bhattacharjee, Amrita and Jiang, Yuxuan and Chen, Canyu and Wu, Tianhao and others},
  booktitle={Proceedings of the 2025 Conference on Empirical Methods in Natural Language Processing},
  pages={2757--2791},
  year={2025}
}

@article{zeng2025futurex,
  title={Futurex: An advanced live benchmark for llm agents in future prediction},
  author={Zeng, Zhiyuan and Liu, Jiashuo and Chen, Siyuan and He, Tianci and Liao, Yali and Tian, Yixiao and Wang, Jinpeng and Wang, Zaiyuan and Yang, Yang and Yin, Lingyue and others},
  journal={arXiv preprint arXiv:2508.11987},
  year={2025}
}

@inproceedings{li2024cryptotrade,
  title={CryptoTrade: A reflective LLM-based agent to guide zero-shot cryptocurrency trading},
  author={Li, Yuan and Luo, Bingqiao and Wang, Qian and Chen, Nuo and Liu, Xu and He, Bingsheng},
  booktitle={Proceedings of the 2024 Conference on Empirical Methods in Natural Language Processing},
  pages={1094--1106},
  year={2024}
}

@misc{caiba2025cryptonerbenchmark,
  title={Crypto Named Entity Recognition Benchmark v0.1},
  author={{CAIBA}},
  year={2025},
  howpublished={\url{https://www.caiba.ai/blogs/5}},
  note={CAIBA blog post, accessed March 26, 2026}
}

@misc{caiba2025onchainexecutionbenchmark,
  title={Onchain Execution Benchmark V0.1},
  author={{CAIBA}},
  year={2025},
  howpublished={\url{https://www.caiba.ai/blogs/6}},
  note={CAIBA blog post, accessed March 26, 2026}
}

@article{mateega2025financeqa,
  title={FinanceQA: a benchmark for evaluating financial analysis capabilities of large language models},
  author={Mateega, Spencer and Georgescu, Carlos and Tang, Danny},
  journal={arXiv preprint arXiv:2501.18062},
  year={2025}
}

@misc{lu2025bizfinbench,
  title={BizFinBench: A Business-Driven Real-World Financial Benchmark for Evaluating LLMs}, 
  author={Guilong Lu and Xuntao Guo and Rongjunchen Zhang and Wenqiao Zhu and Ji Liu},
  year={2025},
  eprint={2505.19457},
  archivePrefix={arXiv},
  primaryClass={cs.AI},
  url={https://arxiv.org/abs/2505.19457}, 
}

@article{zheng2023judging,
  title={Judging llm-as-a-judge with mt-bench and chatbot arena},
  author={Zheng, Lianmin and Chiang, Wei-Lin and Sheng, Ying and Zhuang, Siyuan and Wu, Zhanghao and Zhuang, Yonghao and Lin, Zi and Li, Zhuohan and Li, Dacheng and Xing, Eric and others},
  journal={Advances in neural information processing systems},
  volume={36},
  pages={46595--46623},
  year={2023}
}

@article{chiang2024chatbot,
  title={Chatbot arena: An open platform for evaluating llms by human preference, 2024},
  author={Chiang, Wei-Lin and Zheng, Lianmin and Sheng, Ying and Angelopoulos, Anastasios Nikolas and Li, Tianle and Li, Dacheng and Zhang, Hao and Zhu, Banghua and Jordan, Michael and Gonzalez, Joseph E and others},
  journal={URL https://arxiv. org/abs/2403.04132},
  volume={2},
  number={10},
  year={2024}
}

@misc{li2023alpacaeval,
  title={AlpacaEval: An Automatic Evaluator of Instruction-Following Models},
  author={Li, Xuechen and Zhang, Tianyi and Dubois, Yann and Taori, Rohan and Gulrajani, Ishaan and Guestrin, Carlos and Liang, Percy and Hashimoto, Tatsunori B.},
  year={2023},
  howpublished={\url{https://github.com/tatsu-lab/alpaca_eval}},
  note={LLM-based automatic evaluation benchmark}
}

@inproceedings{chen2024t,
  title={T-eval: Evaluating the tool utilization capability of large language models step by step},
  author={Chen, Zehui and Du, Weihua and Zhang, Wenwei and Liu, Kuikun and Liu, Jiangning and Zheng, Miao and Zhuo, Jingming and Zhang, Songyang and Lin, Dahua and Chen, Kai and others},
  booktitle={Proceedings of the 62nd Annual Meeting of the Association for Computational Linguistics (Volume 1: Long Papers)},
  pages={9510--9529},
  year={2024}
}

@article{yu2024fincon,
  title={Fincon: A synthesized llm multi-agent system with conceptual verbal reinforcement for enhanced financial decision making},
  author={Yu, Yangyang and Yao, Zhiyuan and Li, Haohang and Deng, Zhiyang and Jiang, Yuechen and Cao, Yupeng and Chen, Zhi and Suchow, Jordan W and Cui, Zhenyu and Liu, Rong and others},
  journal={Advances in Neural Information Processing Systems},
  volume={37},
  pages={137010--137045},
  year={2024}
}

@misc{together2025futurebench,
  title={Back to the Future: Evaluating AI Agents on Predicting Future Events},
  author={{Together AI}},
  year={2025},
  howpublished={\url{https://www.together.ai/blog/futurebench}},
  note={FutureBench benchmark for evaluating AI agents on forecasting future events}
}

@article{arora2026predictionmarketbench,
  title={PredictionMarketBench: A SWE-bench-Style Framework for Backtesting Trading Agents on Prediction Markets},
  author={Arora, Avi and Malpani, Ritesh},
  journal={arXiv preprint arXiv:2602.00133},
  year={2026}
}

@article{ye2024mirai,
  title={Mirai: Evaluating llm agents for event forecasting},
  author={Ye, Chenchen and Hu, Ziniu and Deng, Yihe and Huang, Zijie and Ma, Mingyu Derek and Zhu, Yanqiao and Wang, Wei},
  journal={arXiv preprint arXiv:2407.01231},
  year={2024}
}

@article{yu2025livetradebench,
  title={Livetradebench: Seeking real-world alpha with large language models},
  author={Yu, Haofei and Li, Fenghai and You, Jiaxuan},
  journal={arXiv preprint arXiv:2511.03628},
  year={2025}
}

@misc{openai_models_2025,
  title = {OpenAI Models},
  author = {OpenAI},
  year = {2025},
  howpublished = {\url{https://platform.openai.com/docs/models}}
}
